\documentclass[conference]{IEEEtran}
\usepackage{graphicx}
\usepackage{dblfloatfix}
\begin{document}
\title{Visualization Requirements for Business Intelligence Analytics: A Goal-Based, Iterative Framework}

\author{\IEEEauthorblockN{Ana Lavalle, Alejandro Mat{\'e}, Juan Trujillo}
\IEEEauthorblockA{Lucentia Research, DLSI, University of Alicante\\
 Carretera San Vicente del Raspeig s/n, 03690\\
 San Vicente del Raspeig, Alicante, Spain \\ 
 alavalle@dlsi.ua.es, amate@dlsi.ua.es, jtrujillo@dlsi.ua.es}
\and
\IEEEauthorblockN{Stefano Rizzi}
\IEEEauthorblockA{DISI, University of Bologna\\
V.le Risorgimento 2, 40136, Bologna, Italy\\
 stefano.rizzi@unibo.it}}

\maketitle    

\begin{abstract}
Information visualization plays a key role in business intelligence analytics. With ever larger amounts of data that need to be interpreted, using the right visualizations is crucial in order to understand the underlying patterns and results obtained by analysis algorithms. Despite its importance, defining the right visualization is still a challenging task. Business users are rarely experts in information visualization, and they may not exactly know the most adequate visualization tools or patterns for their goals. Consequently, misinterpreted graphs and wrong results can be obtained, leading to missed opportunities and significant losses for companies. The main problem underneath is a lack of tools and methodologies that allow non-expert users to define their visualization and data analysis goals in business terms. In order to tackle this problem, we present an iterative goal-oriented approach based on the i* language for the automatic derivation of data visualizations. Our approach links non-expert user requirements to the data to be analyzed, choosing the most suited visualization techniques in a semi-automatic way. The great advantage of our proposal is that we provide non-expert users with the best suited visualizations according to their information needs and their data with little effort and without requiring expertise in information visualization.

\end{abstract}

\section{Introduction}
Data visualization plays a key role in business intelligence analytics. With ever larger amounts of data that need be interpreted, finding effective visualizations is key to understanding the underlying patterns and the results obtained by analysis algorithms. Without this understanding, users are more likely to distrust the results, following their gut feeling instead of making well-informed decisions. 
Indeed, according to a survey by Salesforce \cite{salesforce2015}, 73\% of high performers strongly agree that analytic tools are valuable for gaining strategic insights from the data.
A large number of companies and researchers are very interested in its application.

Despite this interest, finding the right visualization is still a challenging task. Business users are rarely expert in data visualization, and they may not exactly know what type of information they want to extract from data or which would be the best visualization type. Consequently, misinterpreted graphs and wrong results can be obtained, leading to missed opportunities and significant losses for companies. Another relevant point to be considered is related to dashboard design. A dashboard is a visualization tool that groups multiple tables and charts, ideally aiming to provide a $360^\circ$ view of the phenomenon being analyzed. Dashboards play a key role in the analysis and visualization of data because they enable users --even those with limited ICT skills-- to get their insights and make informed decisions. Although predefined dashboards have been designed for specific sectors, each business and each user may have particular needs different from those already included in predefined dashboards. To design a dashboard, users should state their goals and precisely delimit the information to be represented. However, in most cases, users do not have a clear idea of the most effective visualization techniques for each piece of data.

Although some studies have proposed models to automatically generate dashboards (e.g., \cite{vazquez2018application,santos2017data,kintz2017creating}) they do not consider the best visualization types and tools for each situation. In this direction, some approaches have been proposed to automate data visualization from user requirements (e.g., \cite{borner2014atlas,madhuiba,pena2016exploring,golfarelli2019goal,ehsan2018efficient}). However, these approaches do not \emph{guide} users in the discovery of their objectives nor in the definition of the necessary requirements to generate the most appropriate visualizations for each situation; indeed, they still require users to explicitly state what they wish to visualize and, most importantly, how exactly they want to visualize it.

\begin{figure*}[bp!]
\centering
\includegraphics[width=1\textwidth]{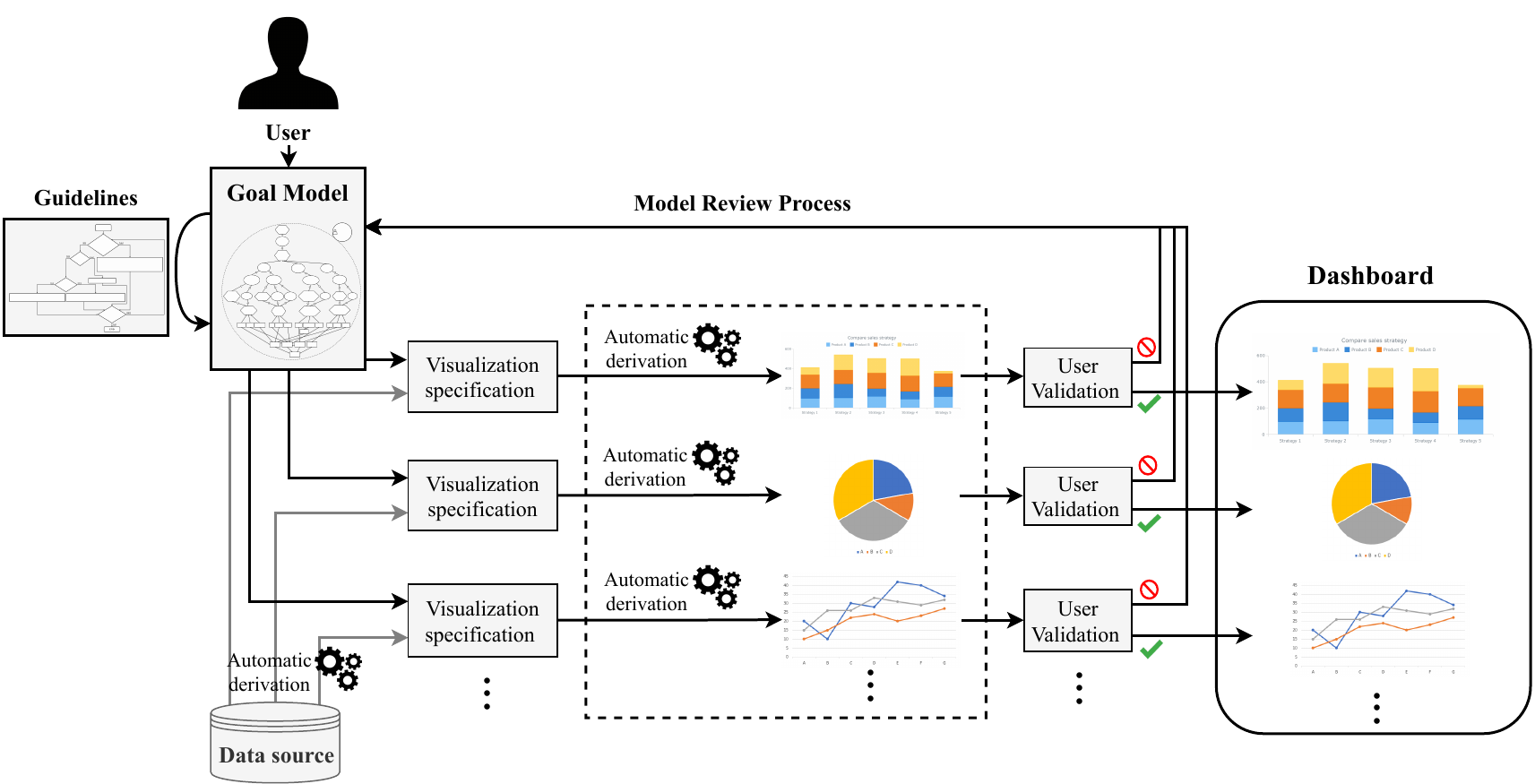}
\caption{Overall view of the process proposed}
\label{fig:process}
\end{figure*}

A very recent approach for the derivation of visualization requirements in analytics is SkyViz \cite{golfarelli2019goal}. In SkyViz, first the user specifies her visualization objectives and describes the dataset to be visualized by defining a \emph{visualization context} based on seven prioritizable visualization requirements. Then, this visualization context is automatically translated into a set of most-suitable visualization types (e.g., pie chart and bar chart) via a skyline-based technique.

As recognized in \cite{golfarelli2019goal}, defining a visualization context from scratch may indeed be a challenge for non-expert users. 
So, in this paper, we complement SkyViz by defining a goal-based \cite{bresciani2004tropos,dalpiaz2016istar,mate2014adding} modeling approach and a set of guidelines to capture user goals and derive the corresponding visualization contexts. Our proposal is meant to work on top of SkyViz making it better usable by non-expert users. Specifically, it improves SkyViz from several points of view: (i) it provides a sequence of steps and guidelines to help users define their goals and achieve them by using the available data sources; (ii) it translates the user's goals into a visualization context; (iii) it semi-automatically extracts visualization requirements from the data sources to be analyzed; and (iv) it provides a rationale for dashboard design. In this way, business users can stay focused on their analysis goals and they can eventually obtain, with a limited effort, the visualizations that best suit their needs. Besides, these visualizations will be grouped into dashboards to allow users to effectively monitor and measure their goals.

Fig. \ref{fig:process} summarizes the process followed in our proposal. In current practice the user is accompanied by a data analyst to help her to follow the process. We are continuing the work on \cite{lavalle2019requirements} where we propose a visualization model for representing visualization details regardless of their implementation technology with the aim to develop a web tool and let users follow the process on its own. In this case, firstly, a sequence of questions guides the user in creating a Goal-Based model that captures her needs. This model encompasses all the visualizations required to tackle the user's objectives. Then, this Goal-Based model is completed by analyzing the features of the data sources to be visualized. At this stage, the model is translated into a a set of visualization contexts, which are then handed to SkyViz to find the best visualization types (process represented within the dashed lines). Finally, each generated visualization is validated by the user to verify if it fulfills the essential requirements for which it was created. The validation process is performed through a questionnaire that is automatically generated from the Goal-Based model, asking the user if the visualization obtained does contribute to answer her goals. Each visualization validated is added to the dashboard. An unsuccessful validation points out to the existence of missing or wrongly-defined requirements that must be reviewed; in this case, a new cycle is started by reviewing the existing model to identify which aspects were not taken into account, generating in turn an updated model. This process is repeated until all user requirements are fulfilled. 

The rest of the paper is structured as follows. Section 2 presents the related work in this area. Section 3 presents our Goal-Based modeling approach for data visualization. Section 4 describes the implementation of our approach. Section 5 discusses an illustrative example in the fiscal domain. Section 6 presents limitations and validity threats of our approach. Finally, Section 7 summarizes the conclusions and our future work.

\section{Related Work}

Several works are focused on finding ways to automatically generate visualizations or dashboards. In \cite{vazquez2018application}, the authors propose an automatic dashboard generator with the capacity to alter dashboard design and functionality without requiring significant development time. In \cite{santos2017data}, a technique is proposed that allows users to modify or add new visualizations as desired, including filters in real time. In \cite{kintz2017creating}, a users-and-roles model is introduced, enabling the automatic generation of user-specific monitoring dashboards, properly displaying the information needed by each user in an organization. All these approaches require that the final user chooses the type of visualization for the representation of the data, without trying to determine which is the most adequate one for the current context. Clearly, this requires the user to be an expert or at least knowledgeable in data visualization techniques.

In order to tackle this problem, some works have proposed different ways to find the best visualization for each analysis. \cite{borner2014atlas} surveys the main classifications proposed in the literature and integrates them into a single framework based on six visualization requirements. In \cite{madhuiba}, authors propose a framework for choosing the best visualization where the main types of charts are related to users goals and to the data dimensionality, cardinality, and the type they support. Finally, \cite{pena2016exploring} proposes a more detailed classification of data types and relates each common type of chart to the users goals it is most compliant with. 

In \cite{golfarelli2019goal} the authors propose SkyViz, an approach to automate the translation of a structured visualization context specified by the user into a suitable visualization. A visualization context consists of seven coordinates, namely goal, interaction, user skills, dimensionality, cardinality, type of the independent variables, and type of the dependent variables.  
Furthermore, in \cite{ehsan2018efficient} a novel utility function and a suite of search schemes for recommending top-k aggregate data visualizations is presented. The utility function recognizes the impact of numerical dimensions on visualization, which is captured by means of multiple objectives, namely, deviation, accuracy, and usability.

Other works are focused on additional issues related to visualization. In
\cite{gray2017understanding} it is argued that one of the reasons for the lack of advanced visualizations are users, who do not often know how they may represent their data.
In \cite{bresciani2015pitfalls} the authors propose a classification of causes of pitfalls, the designer or the user, and they list three types of (negative) effects: \textit{cognitive}, \textit{emotional}, and \textit{social}. More specifically, they state that the cause of a visualization problem can be twofold: the \textit{encoding} (that is, caused by the designer/developer) or the \textit{decoding} (that is, caused by the reader/user). In the latter case, the user who reads the visualization makes a mistake in the interpretation.

Other works \cite{pang1997approaches} also point out that the rendering process introduces uncertainty in all three areas: from the \textit{data collection process}, \textit{algorithmic errors}, and \textit{computational accuracy and precision}. In addition, others like \cite{johnson2003next} have started thinking about visual representations of errors and uncertainties; possible sources of uncertainty are \textit{acquisition} (instrument measurement error, numerical analysis error, statistical variation), \textit{model} (both mathematical and geometric), \textit{transformation} (errors introduced from resampling, filtering, quantization, and rescaling), and \textit{visualization}. 

While certainly adding value to visualizations, these researches focus on the potential pitfalls of blindly using visualization methods without fully understanding the limitations and assumptions of each method and the rationale behind visualizations. In this sense, visualizations should consider the evolving needs of users, taking into account high-level semantics, reasoning about unstructured and structured data, and providing a simplified access and better understanding of data \cite{aufaure2013s}. As such, although often overlooked when designing visualizations, requirement modelling is an important activity \cite{quartel2009goal}, that compensates the little or no attention often paid to (explicitly) representing the reasons, i.e., the \emph{why}, in terms of motivations, rationale, goals, and requirements. This is specially true for goal-based modeling approaches, where the motivations become first-class citizens in the models. 

It is very important that users understand what they are visualizing and why this visualization contributes to reach a goal. Visualizations must be precise and understandable to users to minimize the interpretation mistakes made by both users and designers. In this sense, \cite{akhigbe2017exploiting} shows how IBM Watson Analytics can be used to visualize and analyze data derived from goal-based conceptual models of regulations and regulatory initiatives.

To sum up, none of the approaches summarized above provides a methodology that guides non-expert users in specifying the most adequate set of visualizations and facilitates their implementation into dashboards to be used for data analysis. To bridge the gap between user needs and visualization, goal-based modeling approaches ---which we apply this paper--- emerge as a natural solution.

\section{A Goal-Based Modeling Approach for Visualization Requirements}

\begin{figure*}[t]
\centering
\includegraphics[width=0.97\textwidth]{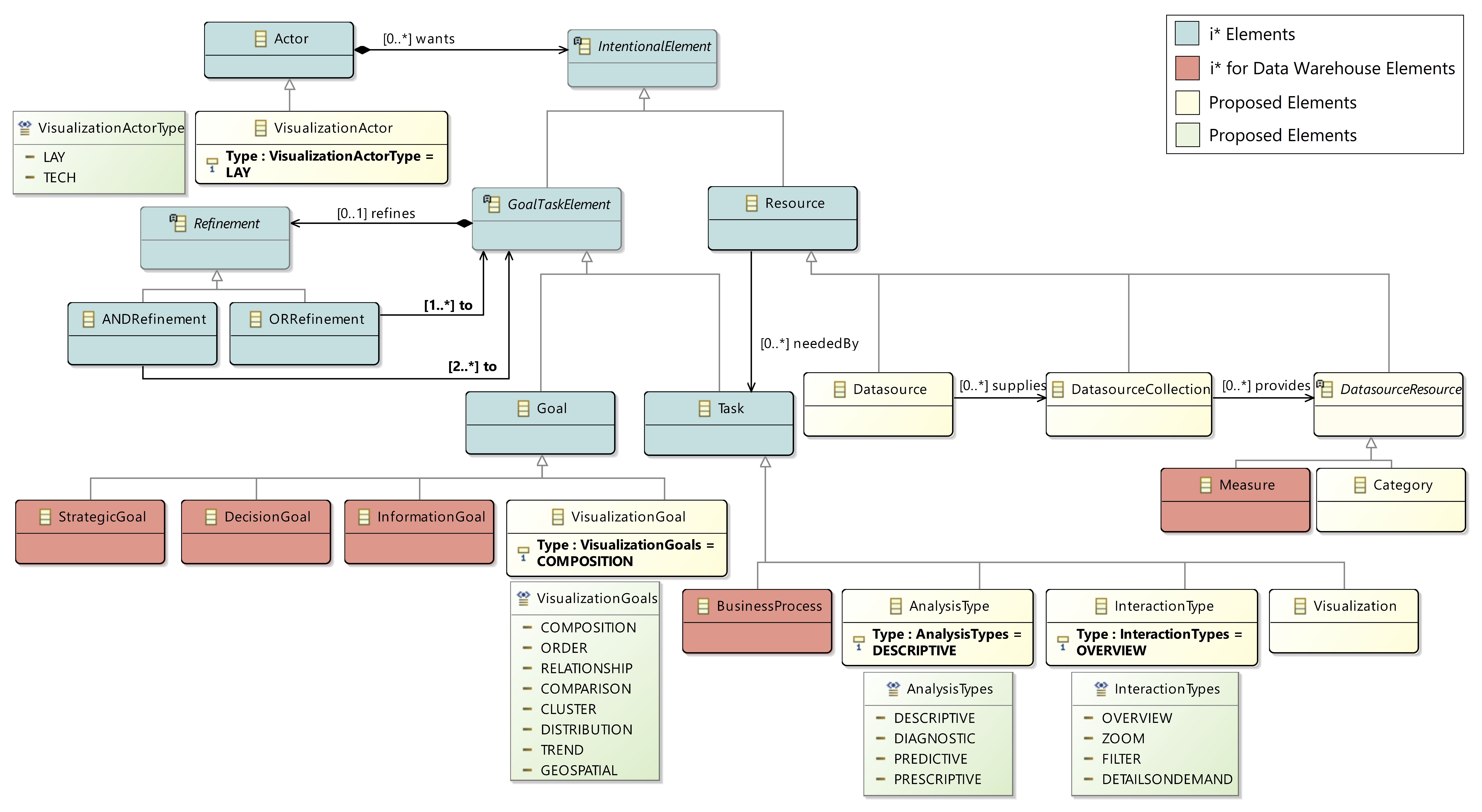}
\caption{Visualization Specification Metamodel}
\label{fig:metamodel}
\end{figure*}

Eliciting from users visualization requirements is considered to be a challenging task, which we aim at supporting in this paper. To this end, we use a combination of modeling and automatic derivation. Initially, we create a Goal-Based model that guides non-expert users towards the specific visualizations they need according to their data analysis objectives. Then, starting from this model we semi-automatically derive a visualization context to be fed into SkyViz and it will recommend us the most suitable visualization type.

The approach we take to formally define our model is through a metamodel that follows the specification given in \cite{golfarelli2019goal} in terms of the coordinates required to build a visualization context (\textit{Goal}, \textit{Interaction}, \textit{User}, \textit{Dimensionality}, \textit{Cardinality}, \textit{Independent Type}, and \textit{Dependent Type}) and the values these coordinates can take.
Our metamodel is shown in Fig. \ref{fig:metamodel} and is an extension of the one used for social and business intelligence modeling, namely i* in its 2.0 version \cite{dalpiaz2016istar} and its \emph{i* for Data Warehouses} extension \cite{mate2014adding}. Existing elements in the i* core are represented in cyan , whereas those included in i* for Data Warehouses are represented in red. The new concepts added by our proposal are represented in light green and yellow. In the following, we describe the concepts included in the metamodel by following the process required for its application.

The aim of the proposed metamodel is to support users in better understanding their objectives and in determining which visualization type they need. To this end, the first element is the \textit{VisualizationActor}, which models the user of the system. There are two types of Visualization Actors: Lay, if she has no knowledge of complex visualizations, and Tech, if she has previous experience and is accustomed to business intelligence analytics.

Once the actor has been defined, the next elements to be defined are the \textit{BusinessProcesses} on which users will focus their analysis. The business process will serve as the guideline for the definition of \textit{Goals}. A goal represents a desired state of affairs with reference to the business process at hand. Goals can be divided into \textit{Strategic}, \textit{Decision}, \textit{Information}, and \textit{Visualization}.

The top-level goals are \textit{StrategicGoals}. They are the main objectives of the business process and are meant as changes from a current situation into a better one. Strategic goals are achieved by means of analyses that support the decision-making process. 

The \textit{AnalysisType} allows users to express which kind of analysis they wish to perform. The definition of a type of analysis will also give the advantage of determining the visualizations to be grouped in the same dashboard. The type of analysis can be determined by selecting which question from the following ones is to be answered \cite{shi2017data}: 

\begin{itemize}
\item \textbf{Prescriptive}: How to act?
\item \textbf{Diagnostic}: Why has this happened?
\item \textbf{Predictive}: What is going to happen?
\item \textbf{Descriptive}: What to do to make it happen?
\end{itemize}

Once the types of analysis to be performed over the strategic goals have been defined, the next elements are \textit{DecisionGoals} and \textit{InformationGoals}. A \textit{DecisionGoal} aims to take appropriate actions to fulfill a strategic goal and explains how it can be achieved. \textit{DecisionGoals} communicate the rationale followed by the decision-making process; however, by themselves they do not provide the necessary details about the data to be visualized. Therefore, for each decision goal there are one or more \textit{InformationGoals}, i.e., lower-level abstraction goals representing the information to be analyzed.

For each \textit{InformationGoals} there will be one \textit{Visualization}. \textit{Visualization} is defined as a task because we understand it as the visualization process, not as the visualization representation. A \textit{Visualization} is characterized by one or more \textit{VisualizationGoals} which describe which aspects of the data the visualization is trying to reflect, and one or more kinds of \textit{InteractionType} that users will need to have with the visualization. \textit{VisualizationGoal} can be defined as Composition, Order, Relationship, Comparison, Cluster, Distribution, Trend, or Geospatial, while \textit{InteractionType} can be Overview, Zoom, Filter, or Details-on-demand \cite{golfarelli2019goal}. Moreover, a \textit{Visualization} will make use of one or more \textit{DatasourceResource} elements to get the relevant data from the data source. 

In the following subsection, we describe in detail the visualization specification process we propose.

\subsection{Visualization Specification}

\begin{figure}[t]
\centering
\includegraphics[width=0.37\textwidth]{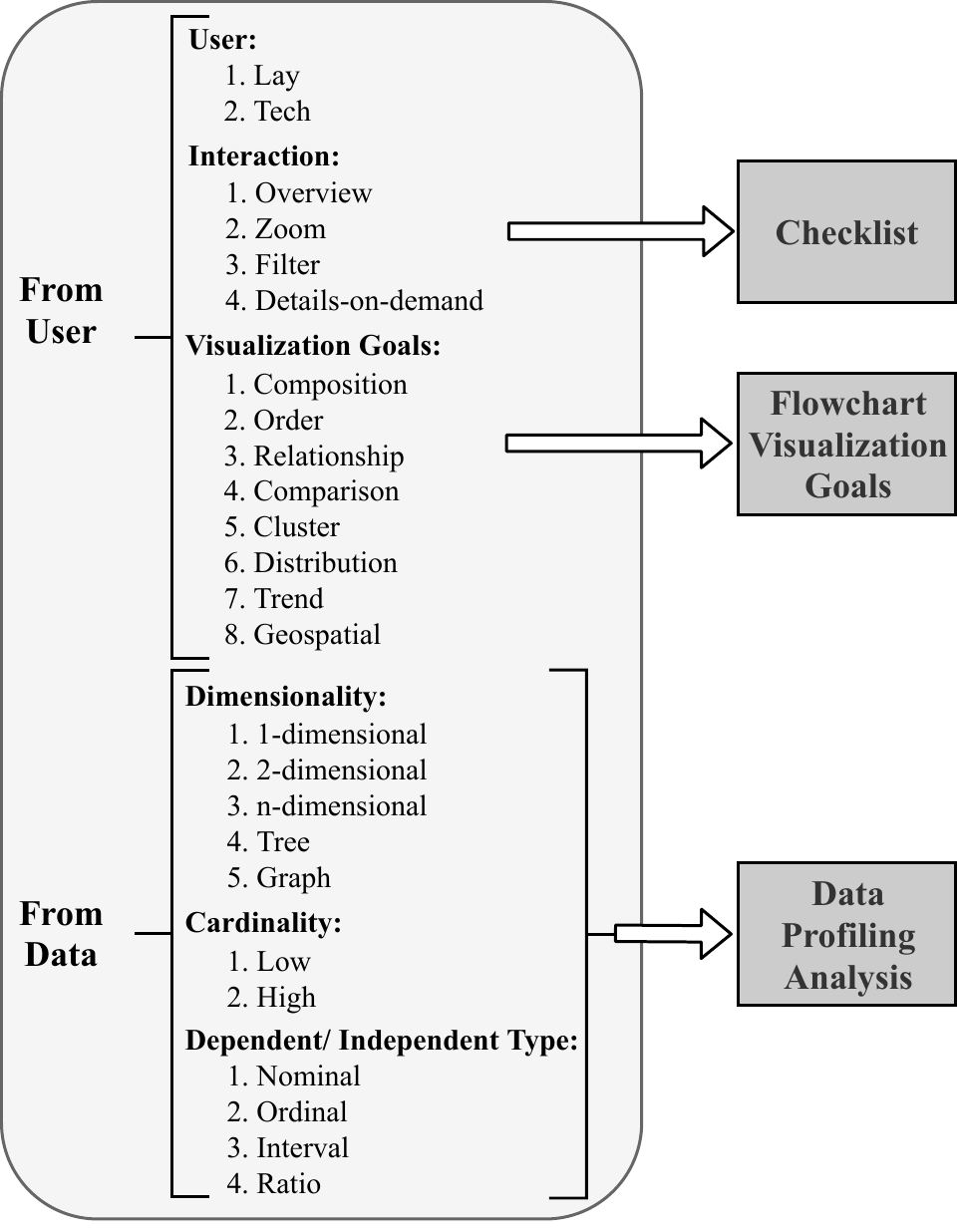}
\caption{Specification of Visualization context}
\label{fig:VisSpe}
\end{figure}

As argued in \cite{golfarelli2019goal}, inexperienced users may find it difficult to properly give values to the seven coordinates included in a visualization context. To facilitate their task, we observe that the coordinates can be split into two families (Fig. \ref{fig:VisSpe}): \emph{user-related} (namely, \textit{Goal}, \textit{Interaction}, \textit{User}) and \emph{data-related} (namely, \textit{Dimensionality}, \textit{Cardinality}, \textit{Independent Type}, and \textit{Dependent Type}). In current practice the user is accompanied by a data analyst to help her to follow the process. We are continuing the work on \cite{lavalle2019requirements} where we propose a visualization model for representing visualization details regardless of their implementation technology with the aim to develop a web tool and let users follow the process on its own.

Therefore, the first step we take concerns user-related coordinates, and consists in guiding users to specify what \textit{Visualization Goals} they aim to achieve and which kind of \textit{Interaction} they would like to have.

\begin{figure}[h!]
\centering
\includegraphics[width=0.42\textwidth]{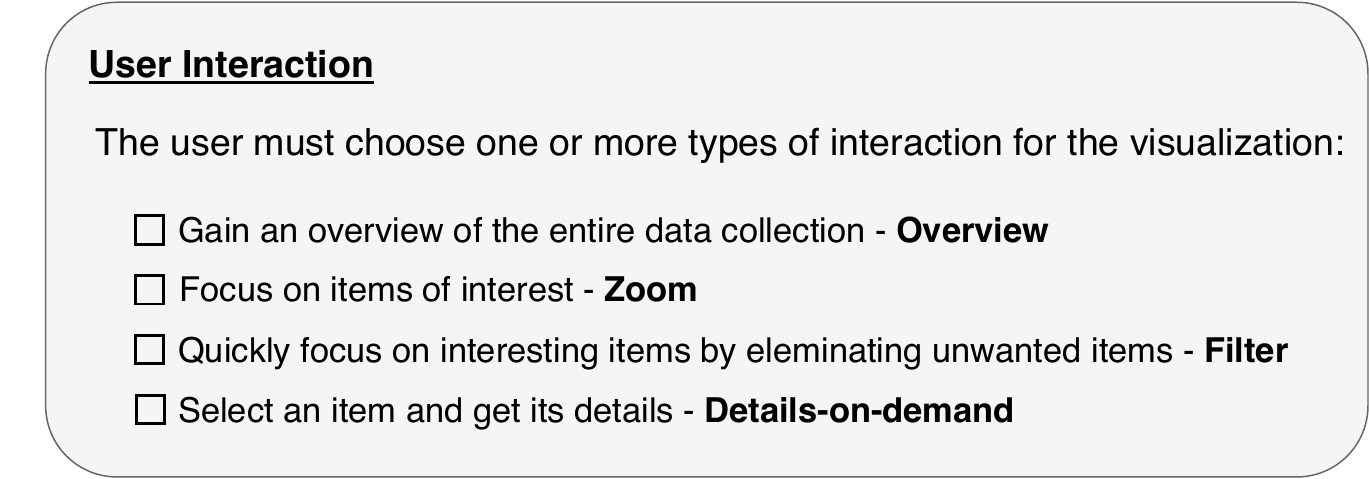}
\caption{Interaction with the visualization}
\label{fig:inte}
\end{figure}

\begin{itemize}
\item \textbf{Interaction:}

The possible interactions are \textit{Overview} (gain an overview of the entire data collection), \textit{Zoom} (focus on items of interest), \textit{Filter} (quickly focus on interesting items by eliminating unwanted items), and \textit{Details-on-demand} (select an item and get its details). We show a checklist to the user (Fig. \ref{fig:inte}) from which she can choose one or more types of interaction. 
\end{itemize}

\begin{figure*}[t]
\centering
\includegraphics[width=0.76 \textwidth]{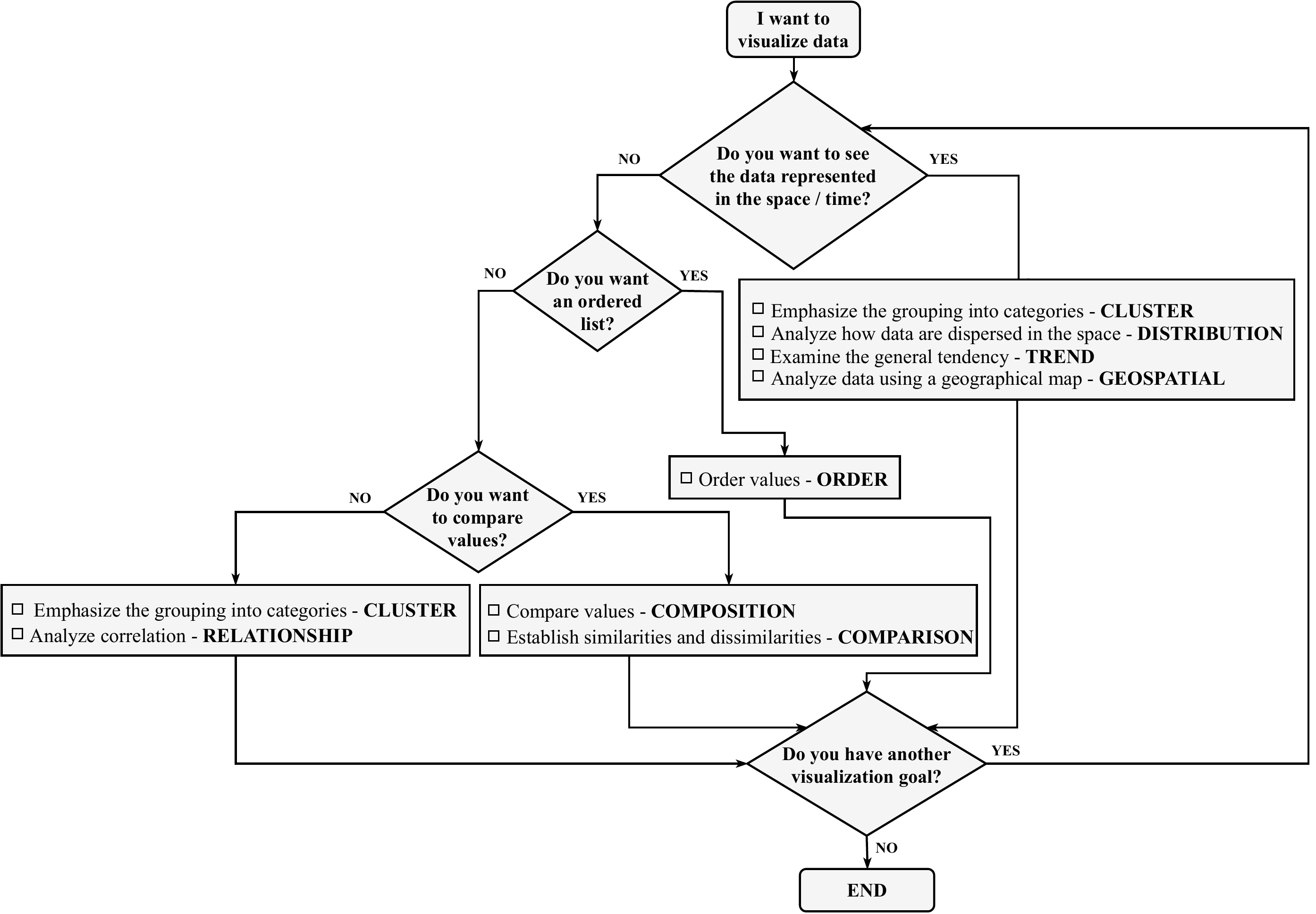}
\caption{Guidelines expressed as a flowchart to help non-expert users in defining visualization goals}
\label{fig:flowchart}
\end{figure*}

\begin{itemize}
\item \textbf{Visualization Goals:}

A visualization goal can be \textit{Composition}, \textit{Order}, \textit{Relationship}, \textit{Comparison}, \textit{Cluster}, \textit{Distribution}, \textit{Trend}, or \textit{Geospatial}. Since choosing the right goal can be difficult depending on the context, to aid users in finding which visualization goal they are pursuing we use the flowchart in Fig. \ref{fig:flowchart}, which contains a series of Yes/No questions to be answered by users. The flowchart provides an easy way to discern which visualization goals should be included for each visualization, thus simplifying the task for non-expert users.
\end{itemize}

As to data-related coordinates, we semi-automatically extract their values by analyzing the features of the data sources. In this way, users do not need to manually inspect the data or have a deep understanding of their characteristics to obtain the most adequate visualizations, and we avoid the introduction of errors in the process.

\begin{itemize}
\item \textbf{Data Profiling Analysis:}

In addition to the requirements provided by users, we extract the values of the remaining coordinates by analyzing the features of the data sources. Users need only to provide the source dataset; then, \textbf{Dimensionality}, \textbf{Cardinality}, and \textbf{Dependent/Independent Type} will be extracted as explained below.

First, users specify a connection to the source dataset they wish to visualize. A menu is provided where users can choose if they want to know the Data type, Cardinality or Dimensionality of the selected column. Finally, this software returns the information requested by users. This development has been created to collect information about the data in a simple way for users. To know how to delimit the values for each coordinate we have followed the proposed in \cite{golfarelli2019goal}. In this way we classify the \textbf{Dimensionality}, \textbf{Cardinality}, and \textbf{Dependent/Independent Type} as follows:

\begin{itemize}
\item 
\textbf{Dependent/Independent Type} is used to declare the type of each variable. It can be \textit{Nominal} when it is qualitative and each variable is assigned to one category, \textit{Ordinal} when it is qualitative and categories can be sorted, \textit{Interval} when it is quantitative and equality of intervals can be determined, or \textit{Ratio} when it is quantitative with a unique and non-arbitrary zero point. 

We delimited each category as follows: If the value is a number, we determine \textit{Ratio} if is a numeric with a unique and non-arbitrary zero point or \textit{Interval} if is a numeric with under 0 values. In the cases where the value is a string of characters, the program shows a grouped list of the values. Then the user is available to determine if in the list there is an order, then it would be \textit{Ordinal}, and if the user can not determine an order it would be \textit{Nominal}.
\end{itemize} 

\begin{itemize}
\item 
\textbf{Cardinality} represents the cardinality of the data, and it can be defined as \textit{Low} or \textit{High} depending of the numbers of items to represent. It will be \textit{Low} cardinality from a few items to a few dozens items and \textit{High} cardinality if there are some dozens items or more. Some visualization types support a larger number of items than others (for example, a pie chart can only visualize low-cardinality data, while a heat map is also fit for high-cardinality data).
\end{itemize} 

\begin{itemize}
\item
\textbf{Dimensionality} is used to declare the number of variables to be visualized. Specifically, it can be \textit{1-dimensional} when the data to represent is a single numerical value or string, \textit{2-dimensional} when one variable depends on other, \textit{n-dimensional} when a data object is a point in an n-dimensional space, \textit{Tree} when a collection of items have a link to one other parent item, or \textit{Graph} when a collection of items are linked to arbitrary number of other items.
\end{itemize} 
\end{itemize}

Once all the requirements have been gathered, we can use SkyViz to get the best type of visualization suited for each particular case while taking into account the preferences of users. However, to check if the visualization generated really fulfills the essential requirements for which it was created, a questionnaire is submitted to the user. The questionnaire will be generated automatically from the information specified by users in the model. Specifically, users will be asked if the visualization contributes to answering the \textit{InformationGoal} defined in the model. If the visualization passes the validation, it will be added to the dashboard. Conversely, if it does not pass the validation, a review of the model will be done to know what aspects were not taken into account and thus generate an updated model. This review gives users an assisted path to improve the obtained visualizations and helps them to achieve their goals.

\section{Implementation}

\begin{figure}[bp!]
\centering
\includegraphics[width=0.43\textwidth]{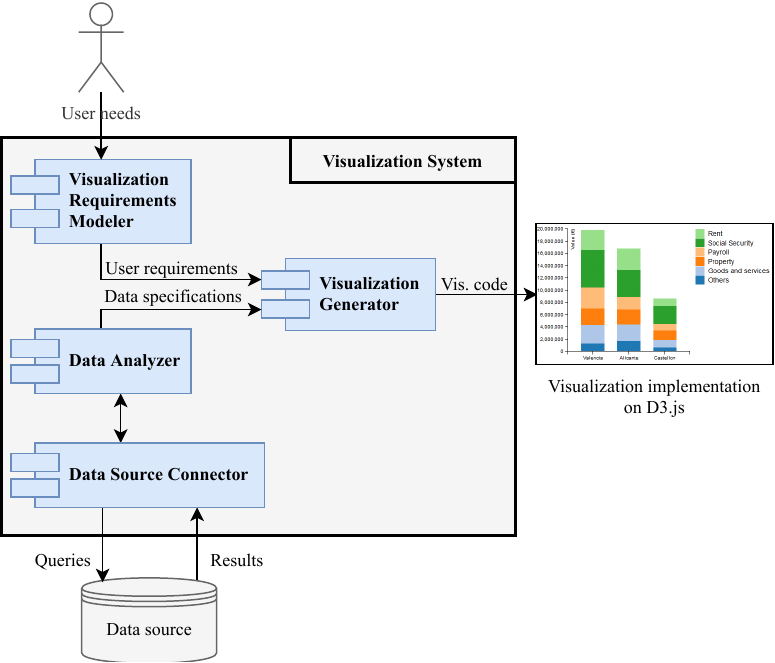}
\caption{System architecture}
\label{fig:archit}
\end{figure}
The implementation of our approach relies on four integrated components as Fig. \ref{fig:archit} shows: (i) the CASE tool aimed at creating the model through the definition of a metamodel, represented as ``Visualization Requirements Modeler''; (ii) ``Data Analyzer'' component that semi-automatically extracts the dataset features, through queries using the (iii) ``Data Source Connector''; (iv) the ``Visualization Generator'', component that selects and renders the best visualization following the process described in \cite{golfarelli2019goal}. These four components are integrated into our system. The system extracts information from users and data sources and it realizes a communication between the components to generate a visualization. 

The CASE tool is implemented in Eclipse by using the Ecore metamodel as a baseline. Defining our metamodel in Ecore enables the automatic generation of the diagram editors for models. Using the Ecore framework we are able to generate the java class objects that support the creation of requirements models. 

The Data Analyzer software created to extract the data profiling has been implemented in Java. It allows users to specify the data source where they need to extract information and performs in an automated and guided way the extraction of information. The MySQL relational database has been used to make the connection, but other types of data sources can be connected as well. In order to use another type of data source we just need to replace the Data Source Connector.

\begin{figure}[h!]
\centering
\includegraphics[width=0.25\textwidth]{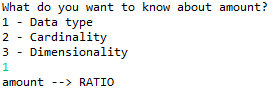}
\caption{Data profiling analysis example}
\label{fig:revIng}
\end{figure}

Fig. \ref{fig:revIng} shows an example of the interactive version of the code that is executed to extract information from the data source. This code will be connected to the user-defined model allowing users to automatically obtains the requested data information.

\begin{figure}[bp!]
\centering
\includegraphics[width=0.35\textwidth]{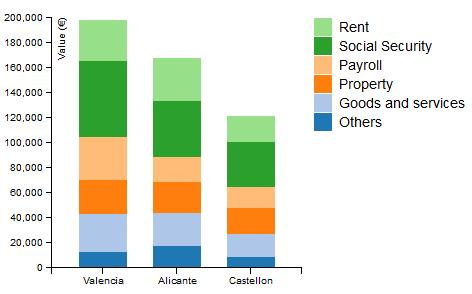}
\caption{Visualization rendered in D3.js}
\label{fig:resultEx}
\end{figure}

Part of the visualization requirements represented in the case tool are elicited from users by following our model, the rest comes from the analysis of the data sources. Once we have all these requirements defined, using SkyViz the visualization context is automatically translated into the most-suitable visualization type. Then, the visual requirements are translated into a call to the D3 JavaScript library \cite{D3js} which renders the visualization. In the cases where a map has to be rendered, the Plotly library \cite{plotly} is used, it can be developed on JavaScript, Python or R. Fig. \ref{fig:resultEx} shows the final result of the process using D3.js. In the next section, we apply our approach to an illustrative example in the field of tax collection.

\section{Illustrative Example}

\begin{figure*}[t]
\centering
\includegraphics[width=0.75\textwidth]{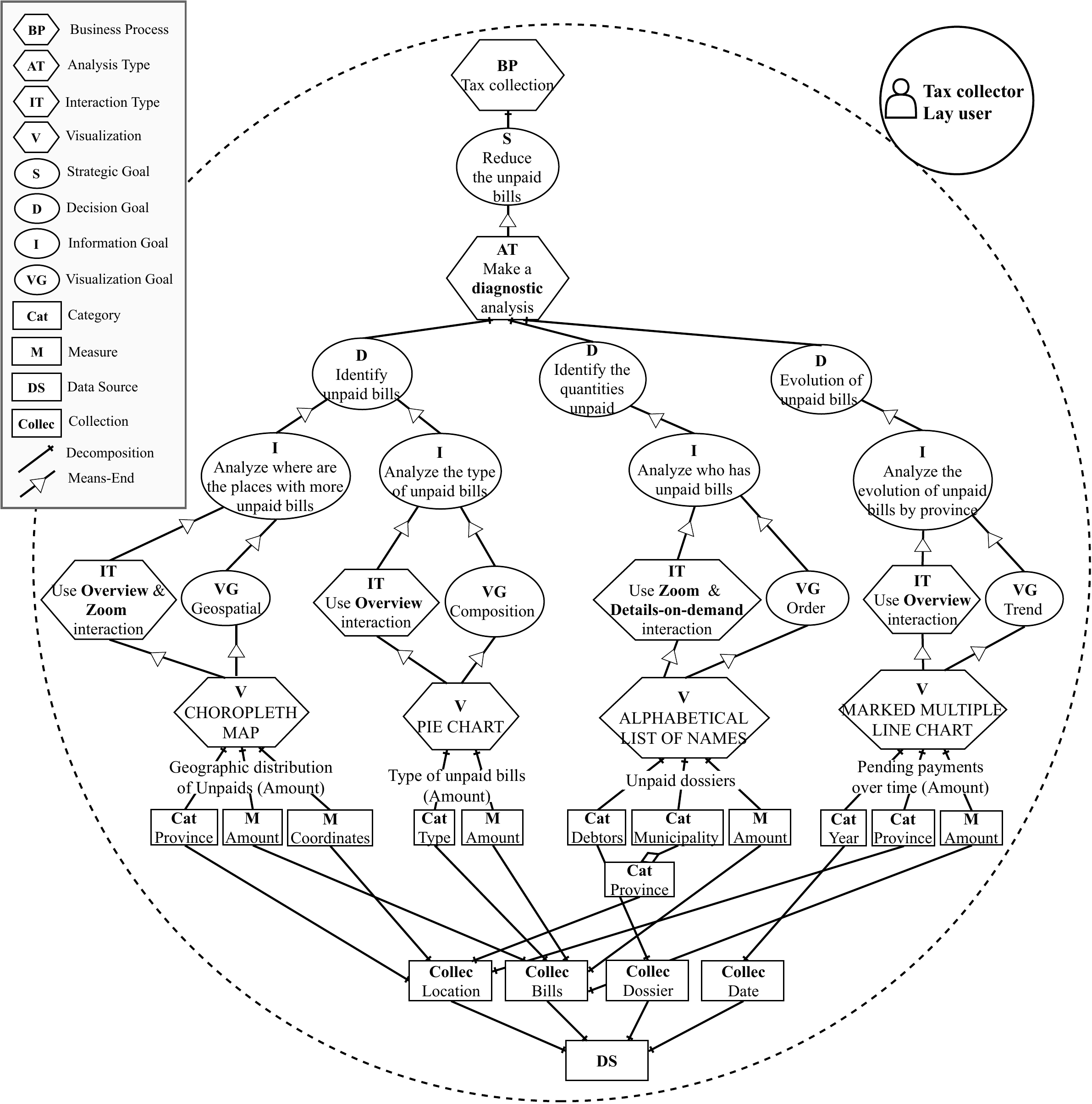}
\caption{Application of our metamodel to the illustrative example}
\label{fig:example}
\end{figure*}

In order to evaluate the validity of our approach we have applied it to an illustrative example. In this case, a tax collection organization has been selected to evaluate the validity of our approach. A tax collection organization requires a set of visualizations to analyze their data in order to help them detect underlying patterns in their unpaid bills and tax collection distribution. Due to the sensibility of their data, we are not allowed to show the real values; besides, data had to be anonymized.

Our approach has been applied to the tax collection organization, producing the model shown in Fig. \ref{fig:example}. In this model, the company wants to analyze the unpaid debts. Therefore, the analysis will focus on the \textit{``Tax collection''} business process. Defining a business process helps determining which concepts are involved in the analysis and what kind of goals are pursued. Here the user is a tax collector who is not a specialist in analytics but rather an expert in tax management, thus she is defined as \textit{``Lay user''}.

\subsection{Specifying Goals}
The main objectives of the business process are defined as shown in Fig. \ref{fig:example}. Specifically, the user defined her strategic goal as \textit{``Reduce the unpaid bills''}. Strategic goals are achieved by means of analyses that support the decision-making process. The analysis type allows users to express what kind of analysis they wish to perform. In this case, 
the user wishes to know why bills are unpaid. Thus, she decides to perform a \textit{``Diagnostic analysis''}. Having defined a specific type of analysis, we are aware that all context information for \textit{``Diagnostic analysis''} should be gathered in the same dashboard in order to provide a complete answer to the user.

The diagnostic analysis is decomposed into decision goals. A decision goal aims to take appropriate actions to fulfill a strategic goal and explains how it can be achieved. The user defined her decisions goals as: \textit{``Identify unpaid bills''}, \textit{``Identify the quantities unpaid''}, and \textit{``Evolution of unpaid bills''}. 
Decisions goals communicate the rationale followed by the decision-making process; however, by themselves they do not provide the necessary details about the data to be visualized. Therefore, for each decision goal we specify one or more information goals, i.e., lower-level abstraction goals representing the information to be analyzed.

From each of the decision goals she listed, the user refined the following information goals: \textit{``Analyze where are the places with more unpaid bills''}, \textit{``Analyze the type of unpaid bills''}, \textit{``Analyze who has unpaid bills''}, and \textit{``Analyze the evolution of unpaid bills by province''}. Information goals represent the lowest level of goal abstraction.

At this point, the user has the necessary information about her goals to start defining the visualization context. For each information goal, we will have one visualization to achieve it. A visualization is characterized by one or more visualization goals which describe what aspects of the data the visualization is trying to reflect, and one or more kinds of interaction that they will like to have with the visualization. Moreover, a visualization will make use of one or more data source elements to get the relevant data from the database. 

In this case, the user defines the interactions she wants to have with each visualization following the checkbox shown in Fig. \ref{fig:inte}. \textit{``Overview''}, \textit{``Zoom''} and \textit{``Details-on-demand''} have been defined. Additionally, following the flowchart shown in Fig. \ref{fig:flowchart}, the user specified her visualization goals: \textit{``Geospatial''}, \textit{``Composition''}, \textit{``Order''}, and \textit{``Trend''}.

\begin{table*}[bp!]
  \begin{center}
    \caption{Suitability scores needed for the two visualization context of the illustrative example.}
    \label{tab:table1}
    \begin{tabular}{r l|c|c|c|}
       & \textbf{VISUALIZATION} & & & \\
      & \textbf{CONTEXT}   & \textbf{Stacked Column Chart} & \textbf{Bubble Chart} & \textbf{Pie Chart}\\
      \hline
      \textbf{Goal:} 
        & Composition & fit & unfit & fit\\
        & Comparison & fit & fit & unfit\\
     \textbf{Interaction:} 
        & Overview & acceptable & acceptable & fit\\
     \textbf{User:} 
        & Lay  & fit & acceptable & fit\\
     \textbf{Dimensionality:} 
        & 2-dimensional   & unfit & unfit & fit\\
        & n-dimensional   & fit & fit & unfit\\
      \textbf{Cardinality:} 
        & Low  & fit & acceptable  & fit\\
      \textbf{Independent Type:} 
        & Nominal  & fit & unfit & fit\\
     \textbf{Dependent Type:} 
        & Ratio & fit & fit & fit\\
    \hline
    \end{tabular}
  \end{center}
\end{table*}

Finally, the visualizations are decomposed into Categories and Measures that will populate them. In this case, the visualization of \textit{``Geographic distribution of unpaids''} includes \textit{``Province''} as category, and \textit{``Amount''} and \textit{``Coordinates''} as measures. These attributes come from the data source collections \textit{``Location''} and \textit{``Bills''}, respectively. For the visualization of \textit{``Type of unpaid bills''} the user picked \textit{``Type''} as relevant category, and \textit{``Amount''} as measure. These are obtained from the data source collection \textit{``Bills''}. Next, in the case of \textit{``Unpaid dossiers''}, categories \textit{``Debtors''}, \textit{``Municipality''}, and \textit{``Province''} as well as measure \textit{``Amount''} are selected  from the data source collections \textit{``Dossier''}, \textit{``Location''} and \textit{``Bills''}. The last visualization is \textit{``Pending payments over time''}, that makes use of categories \textit{``Year''} and \textit{``Province''} and of measure \textit{``Amount''}. These data are obtained from the collections \textit{``Date''}, \textit{``Location''}, and \textit{``Bills''}.

Once user have defined the data sources and collections from where the data will be extracted, it is possible to profile data sources to determine Dimensionality, Cardinality and Dependent/Independent Type.

\subsection{Profiling Data Sources}
Tax data are divided into different collections as follows: \textbf{Location} collects information about where tax was unpaid; \textbf{Date} holds data about when it was unpaid; \textbf{Dossier} represents who is the debtor, whether a person or an entity; and \textbf{Bill} joins the set of previously mentioned data by means of bills. Each collection is futter decomposed into measures and categories. 

Next step is to analyze the data sources in order to extract information about information about Dimensionality, Cardinality, Dependent/Independent Type from the data sources, using our Data Analyzer tool as shown in Section 3.

We focus on the \textbf{``Type of unpaid bills''} visualization from our Goal-Based model, which requires information about the category \textit{``Type''} and measure \textit{``Amount''}. 
Firstly, using the data profiling tool, the independent variable \textit{``Type''} are classified as \textbf{Nominal} and the dependent variable \textit{``Amount''} as \textbf{Ratio}. Dimensionality is set to \textbf{2-dimensional}, because the user has defined 2 variables to visualize. Finally, the tool computes the Cardinality of data through a query. The tool defines Cardinality as \textbf{Low} because the data contains a few items to represent, there are 6 types of bills. Overall, the values obtained through data profiling are: 

\begin{itemize}
\item \textbf{Dimensionality}: 2-dimensional
\item \textbf{Cardinality}: Low
\item \textbf{Independent Type}: Nominal
\item \textbf{Dependent Type}: Ratio
\end{itemize}

\subsection{Validation and Results}

\begin{figure}[bp!]
\centering
\includegraphics[width=0.22\textwidth]{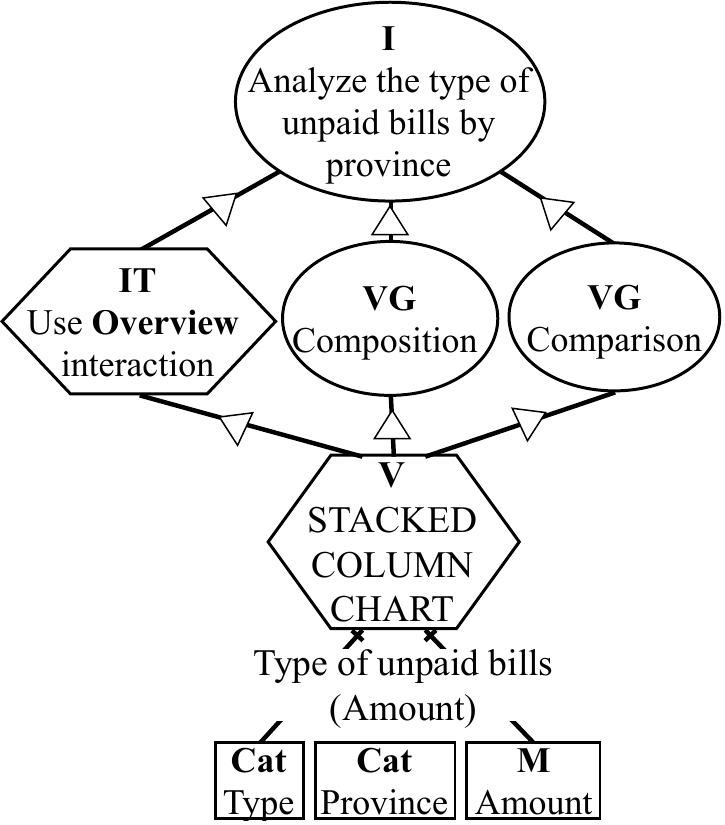}
\caption{Model update due the review}
\label{fig:update}
\end{figure}

Once the visualization has been specified, the visualization context is used as input to be fed into SkyViz and it will recommend us the most suitable visualization type by a table with suitability scores. 

The suitability scores from Table 1 are proposed by \cite{golfarelli2019goal} and shows the visualization context we derived from the information goal \textit{``Analyze the type of unpaid bills''}, together with the suitability scores for tree visualization types, namely, \textit{``Stacked Column Chart''}, \textit{``Bubble Chart''}, and \textit{``Pie Chart''}. The semantics of the suitability scores in this context is as follows:

\begin{itemize}
\item \textbf{Fit}: Means that the visualization type is fully compatible.
\item \textbf{Acceptable}: Means that the visualization type is compatible with the coordinate value, though it may fail to emphasize some of the required features.
\item \textbf{Discouraged}: Means that the visualization type can be used in principle for the coordinate value, but it may distort the very nature of the required features.
\item \textbf{Unfit}: Means that the visualization type should not be used for the coordinate value.
\end{itemize}

\begin{figure*}[bp!]
\centering
\includegraphics[width=0.55\textwidth]{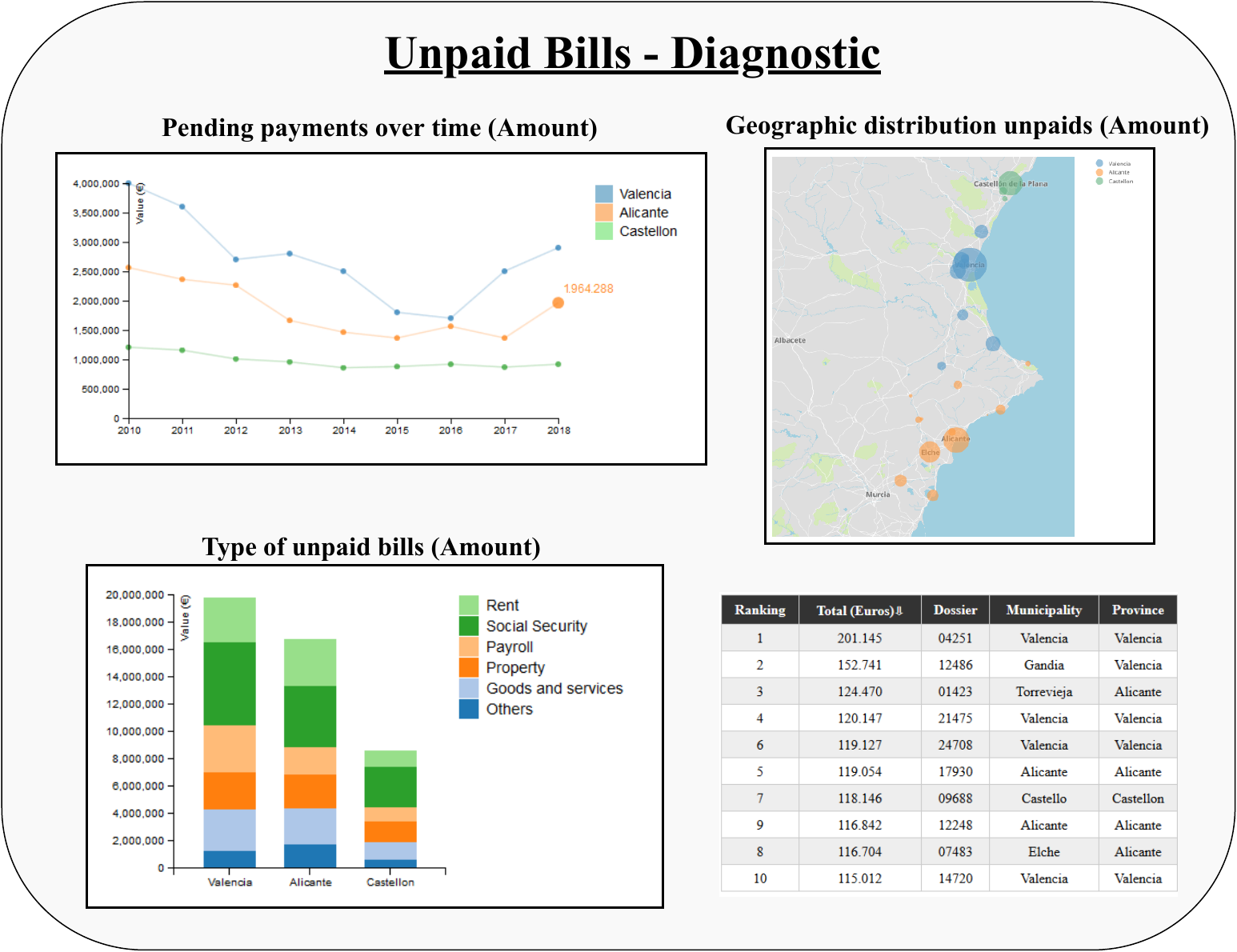}
\caption{Dashboard for tax collection analysis}
\label{fig:result}
\end{figure*}

Accordingly the suitability scores in Table 1, the most suitable visualization for our analysis is \textit{``Pie Chart''}. A mockup of the pie chart visualization is shown to the user and the user detects that this visualization does not reach exactly her goals. Consequently, a model review is done by the user and she detects that the information goal \textit{``Analyze the type of unpaid bills''} is not correctly defined, she adds information modifying it to \textit{``Analyze the type of unpaid bills''}. She continues reviewing the model and she extends the visualization goals by adding \textit{``Comparison''}. Finally, she select from the collection \textit{``Location''} the category \textit{``Province''} to be represented in the visualization. This review modifies the model updating it as shown in Fig. \ref{fig:update}. The goals have change to \textit{``Composition''} and \textit{``Comparison''} and dimensionality now is \textit{``n-dimensional''}.

Now, with the update of the context, the most suitable visualization for our visualization context has changed to \textit{``Stacked Column Chart''}. Again a mockup of the visualization is shown to the user and now the visualization can effectively answer the user information goal \textit{``Anayze the type of unpaid bills by province''}. Therefore, following the derivation process we make a call to the D3 JavaScript library, obtaining the visualization. Consequently, the visualization is added to the dashboard, as shown in the lower-left corner of Fig. \ref{fig:result}. 

This visualization, combined with those generated from the informational goals \textit{``Analyze where are the places with more unpaid bills''}, \textit{``Analyze who has unpaid bills''} and \textit{``Analyze the evolution of unpaid bills by provinces''}, are grouped into the dashboard layout proposed in Fig. \ref{fig:result}, aimed at satisfying the analytic requirements of our tax collector user with the most adequate visualizations.

\section{Limitations and Validity Threats}
In this section, we summarize the main limitations we envision for our approach.

\begin{itemize}
\item 
Up to now, we have had satisfactory results in our applications to real cases and testing the proposal with a focus group. However, since we have not yet tested the proposal in a comprehensive set of contexts, it may be the case that some specific user profiles have not yet been identified.
\item
In principle, our proposal is context-independent; up to now, it has been applied in the economic, educational, and gas turbine contexts. However, some other context may raise specific issues that we have not contemplated yet. For instance, some contexts may require visualizations to be produced in real time, which is currently out of the scope of our approach.
\item
In our experiments, we have worked with a data analyst supporting each non-expert user in defining her visualization requirements. We are currently working to conclude the development of the tool proposed in \cite{lavalle2019requirements} to verify that users are actually qualified to define visualization requirements completely on their own.
\item
At the time of defining the visualization specification, the user has to know the features of the dataset to be visualized. 
Besides, she is required to be expert in the application domain for which visualization is required.
\item
We rely on SkyViz to derive the best suited visualizations type, however SkyViz itself has some limitations \cite{golfarelli2019goal}. First of all, it currently includes a limited number of visualization types. On the other hand, if a significantly larger number of visualization types were included, the seven coordinates might no longer be sufficient to distinguish them, in which case the user would be provided with a large number of (probably similar) visualization types. To cope with this situation, other coordinates should be added, but the research questions to be addressed would be (i) how to select them in order to actually improve the discriminatory power of SkyViz, and (ii) how to deal with these new coordinates in the goal-based model.
\item
While our proposal would be able to represent collaborative visualizations through \textit{Strategy Dependency} diagrams, this aspect has not been yet fully explored and is considered out of the scope of this paper.
\end{itemize}

\section{Conclusions and Future Work}

In this paper, we have presented an iterative goal-based modeling approach in order to help non-expert users define their data analysis goals and derive the most adequate visualizations to facilitate the analysis of data. Compared to other approaches, our proposal covers the whole process from the definition and modeling of user requirements to the implementation of the visualizations. The great advantage of our proposal is that non-technical users can effectively communicate their visual analytic needs without needing deep knowledge of visualization technologies or data sources descriptions. Furthermore, visualizations are easily modified by altering requirements such as the type of interaction or the visualization goal pursued.

As part of our future work, we are working on improving the data analysis step to better support the detection of independent and dependent variables when multiple measures and categories are present. Furthermore, we are working on the implementation of a user-friendly diagram editor by using Graphiti. In this way, we will be able to provide better support for users even when there is no analyst available to aid them when building the requirements model. We will also consider capturing non-functional requirements or quality goals to help decide between visualizations.

\section{Acknowledgment}
This work has been co-funded by the ECLIPSE-UA (RTI2018-094283-B-C32) project funded by Spanish Ministry of Science, Innovation, and Universities. Ana Lavalle holds an Industrial PhD Grant (I-PI 03-18) co-funded by the University of Alicante and the Lucentia Lab Spin-off Company.

\bibliographystyle{splncs04}
\bibliography{mybibliography}

\begin{thebibliography}{10}
\providecommand{\url}[1]{\texttt{#1}}
\providecommand{\urlprefix}{URL }
\providecommand{\doi}[1]{https://doi.org/#1}

\bibitem{akhigbe2017exploiting}
Akhigbe, O., Heap, S., Amyot, D., Richards, G.: Exploiting {IBM} watson analytics to visualize and analyze data from goal-based conceptual models. In: Proceedings of the {ER} Forum 2017 and the {ER} 2017 Demo Track co-located with the 36th International Conference on Conceptual Modelling. pp. 338--342 (2017)

\bibitem{lavalle2019requirements}
Ana~Lavalle, A.M., Trujillo, J.: Requirements-driven visualizations for big data analytics: a model-driven approach. In: International Conference on Conceptual Modeling {ER} 2019, to appear. Springer (2019)

\bibitem{aufaure2013s}
Aufaure, M.: What's up in business intelligence? {A} contextual and knowledge-based perspective. In: Conceptual Modeling - 32th International Conference, {ER} 2013. pp. 9--18. Springer (2013)

\bibitem{borner2014atlas}
B{\"o}rner, K.: Atlas of knowledge (2014)

\bibitem{D3js}
Bostock, M.: Data-driven documents (2019), \url{https://d3js.org/}

\bibitem{bresciani2004tropos}
Bresciani, P., Perini, A., Giorgini, P., Giunchiglia, F., Mylopoulos, J.: Tropos: An agent-oriented software development methodology. Autonomous Agents and Multi-Agent Systems  \textbf{8}(3),  203--236 (2004)

\bibitem{bresciani2015pitfalls}
Bresciani, S., Eppler, M.J.: The pitfalls of visual representations: A review and classification of common errors made while designing and interpreting visualizations. Sage Open  \textbf{5}(4),  2158244015611451 (2015)

\bibitem{dalpiaz2016istar}
Dalpiaz, F., Franch, X., Horkoff, J.: istar 2.0 language guide. CoRR  \textbf{abs/1605.07767} (2016)

\bibitem{ehsan2018efficient}
Ehsan, H., Sharaf, M.A., Chrysanthis, P.K.: Efficient recommendation of aggregate data visualizations. {IEEE} Trans. Knowl. Data Eng.  \textbf{30}(2),  263--277 (2018)

\bibitem{golfarelli2019goal}
Golfarelli, M., Rizzi, S.: A model-driven approach to automate data visualization in big data analytics. Information Visualization, to appear  (2019)

\bibitem{gray2017understanding}
Gray, C.C., Teahan, W.J., Perkins, D.: Understanding our analytics: A visualization survey. Journal of Learning Analytics, to appear  (2017)

\bibitem{johnson2003next}
Johnson, C.R., Sanderson, A.R.: A next step: Visualizing errors and uncertainty. {IEEE} Computer Graphics and Applications  \textbf{23}(5),  6--10 (2003)

\bibitem{kintz2017creating}
Kintz, M., Kochanowski, M., Koetter, F.: Creating user-specific business process monitoring dashboards with a model-driven approach. In: Proceedings of the 5th International Conference on Model-Driven Engineering and Software Development, {MODELSWARD}. pp. 353--361 (2017)

\bibitem{madhuiba}
Madhu~Sudhan, S., Chandra, J.: Iba graph selector algorithm for big data visualization using defence dataset. International Journal of Scientific \& Engineering Research  \textbf{4}(3) (2013)

\bibitem{mate2014adding}
Mat{\'e}, A., Trujillo, J., Franch, X.: Adding semantic modules to improve goal-oriented analysis of data warehouses using i-star. Journal of systems and software  \textbf{88},  102--111 (2014)

\bibitem{pang1997approaches}
Pang, A., Wittenbrink, C.M., Lodha, S.K.: Approaches to uncertainty visualization. The Visual Computer  \textbf{13}(8),  370--390 (1997)

\bibitem{pena2016exploring}
Pe{\~{n}}a, O., Aguilera, U., L{\'{o}}pez{-}de{-}Ipi{\~{n}}a, D.: Exploring {LOD} through metadata extraction and data-driven visualizations. Program  \textbf{50}(3),  270--287 (2016)

\bibitem{plotly}
Plotly: Dash (2019), \url{https://plot.ly/}

\bibitem{quartel2009goal}
Quartel, D.A.C., Engelsman, W., Jonkers, H., van Sinderen, M.: A goal-oriented requirements modelling language for enterprise architecture. In: Proceedings of the 13th {IEEE} International Enterprise Distributed Object Computing Conference, {EDOC} 2009. pp. 3--13. IEEE (2009)

\bibitem{salesforce2015}
Salesforce: State of analytics (2015), \url{www.lubbersdejong.nl/wp-content/uploads/2015/ 10/Salesforce-2015-State-of-Analytics-report.pdf}

\bibitem{santos2017data}
Santos, H., Dantas, V., Furtado, V., Pinheiro, P., McGuinness, D.L.: From data to city indicators: {A} knowledge graph for supporting automatic generation of dashboards. In: The Semantic Web - 14th International Conference, {ESWC}. pp. 94--108. Springer (2017)

\bibitem{shi2017data}
Shi-Nash, A., Hardoon, D.R.: Data analytics and predictive analytics in the era of big data. Internet of Things and Data Analytics Handbook pp. 329--345 (2017)

\bibitem{vazquez2018application}
V{\'{a}}zquez{-}Ingelmo, A., Garc{\'{\i}}a{-}Pe{\~{n}}alvo, F.J., Ther{\'{o}}n, R.: Application of domain engineering to generate customized information dashboards. In: Learning and Collaboration Technologies. Learning and Teaching - 5th International Conference, {LCT}. pp. 518--529. Springer (2018)

\end{thebibliography}

\end{document}